\documentclass[aps,prd,twocolumn,floatfix,nofootinbib,showpacs,superscriptaddress]{revtex4-1}

\usepackage{amsmath,amsfonts,amssymb,bm}
\usepackage{dsfont}
\usepackage{graphicx}
\usepackage{color}
\usepackage{soul}

\usepackage[mathscr,scaled=1.15]{urwchancal}
\DeclareFontFamily{OT1}{pzc}{}
\DeclareFontShape{OT1}{pzc}{m}{it}%
{<-> s * [1.15] pzcmi7t}{}
\DeclareMathAlphabet{\mathpzc}{OT1}{pzc}{m}{it}

\definecolor{purple}{rgb}{0.5,0,0.5}
\definecolor{blue}{rgb}{0.0,0,0.9}
\usepackage[hypertex]{hyperref}
\hypersetup{colorlinks=true, linkcolor=purple, citecolor= purple, urlcolor=blue}

\begin{document}

\title{Phase diagram and thermal properties of strong-interaction matter}
\author{Fei Gao}
\affiliation{Department of Physics and State
Key Laboratory of Nuclear Physics and Technology, Peking
University, Beijing 100871, China}

\author{Jing Chen}
\affiliation{Department of Physics and State
Key Laboratory of Nuclear Physics and Technology, Peking
University, Beijing 100871, China}

\author{Yu-Xin Liu}
\email[Corresponding author: ]{yxliu@pku.edu.cn}
\affiliation{Department of Physics and State Key Laboratory of
Nuclear Physics and Technology, Peking University, Beijing 100871,
China}
\affiliation{Collaborative Innovation Center of Quantum Matter, Beijing 100871, China}
\affiliation{Center for High Energy Physics, Peking
University, Beijing 100871, China}

\author{Si-Xue Qin}
\affiliation{Division of Physics, Argonne National Laboratory, Argonne, IL 60439, USA}

\author{Craig D. Roberts}
\email[Corresponding author: ]{cdroberts@anl.gov}
\affiliation{Division of Physics, Argonne National Laboratory, Argonne, IL 60439, USA}

\author{Sebastian M. Schmidt}
\affiliation{Institute for Advanced Simulation, Forschungszentrum J\"ulich and JARA, D-52425 J\"ulich, Germany}

\date{4 April 2016}

\begin{abstract}
We introduce a novel method for computing the $(\mu,T)$-dependent pressure in continuum QCD; and therefrom obtain a complex phase diagram and predictions for thermal properties of the dressed-quark component of the system, providing the in-medium behaviour of the related trace anomaly, speed of sound, latent heat and heat capacity.
\end{abstract}



\maketitle

\section{Introduction}
Strong interactions in the Standard Model are described by quantum chromodynamics (QCD), which is supposed to describe a vast array of phenomena, from gluon and quark interactions at the highest energies achievable with the large hadron collider to the nature of nuclear material at the core of a compact star.  This last challenge initiated the quest to uncover the equation of state (EoS) for superdense nuclear matter \cite{Collins:1974ky}.  The intervening years have seen remarkable activity, highlighted by
the discovery of a new state of matter, \emph{viz}.\ a strongly-coupled quark-gluon plasma (sQGP) \cite{Gyulassy:2004zy}.  These efforts have delivered a sketch of the QCD EoS in the plane spanned by baryon chemical potential ($\mu_B$) and temperature ($T$) \cite{BraunMunzinger:2008tz, Fukushima:2010bq}.

In vacuum, \emph{i.e}.\ in the neighbourhood $\mu_B \simeq 0 \simeq T$, QCD exists in a phase characterised by two emergent phenomena: confinement and dynamical chiral symmetry breaking (DCSB).  Confinement is most simply defined empirically: those degrees-of-freedom used in defining the QCD Lagrangian -- gluons and quarks -- do not exist as asymptotic states.  
The forces responsible for confinement appear to generate more than 98\% of the mass of visible matter \cite{national2012NuclearS}.  This phenomenon is known as DCSB.  It is a quantum field theoretical effect, which is expressed and explained via the appearance of momentum-dependent mass-functions for quarks \cite{Bhagwat:2003vw, Bowman:2005vx, Bhagwat:2006tu, Roberts:2007ji} and gluons \cite{Cornwall:1981zr, Aguilar:2006gr, Boucaud:2011ug, Ayala:2012pb, Binosi:2014aea} even in the absence of any Higgs-like mechanism.

On the other hand, in medium, \emph{i.e}.\ as $\mu_B$ and/or $T$ are increased beyond certain critical values, the property of asymptotic freedom \cite{Politzer:2005kc, Wilczek:2005az, Gross:2005kv} suggests that QCD undergoes phase transitions.  In the new phases, DCSB disappears and/or gluons and quarks are deconfined.  Indeed, the possibility that the transitions are related and coincident in the $(\mu_B,T)$-plane is much discussed.

The QCD EoS and related thermal properties are important for numerous reasons \cite{Rischke:2003mt}, \emph{e.g}.\
they are crucial inputs to the hydrodynamic simulations used to connect basic theoretical predictions with modern experimental data.  In this context, the current sketches are not adequate: a better picture is needed in order to understand the data and hence the qualities of a sQGP.

The $T\neq 0$ properties of QCD have been scrutinised via simulations of the lattice-regularised theory (lQCD).  This is apparent, \emph{e.g}.\ in Refs.\,\cite{Philipsen:2012nu, Braun-Munzinger:2014pya}, which also highlight problems impeding the extension of lattice methods to $\mu_B\neq 0$.  At this stage, a picture of the phase diagram in the entire $(\mu_B,T)$-plane requires other methods, so models continue to be employed widely.  However, they are numerous in number and various in formulation, and too often provide conflicting predictions \cite{Braun-Munzinger:2014pya}.  Herein, therefore, we analyse aspects of the thermal properties of QCD using methods of continuum quantum field theory; namely, the Dyson-Schwinger equations (DSEs) \cite{Roberts:2000aa, Maris:2003vk, Chang:2011vu, Bashir:2012fs, Cloet:2013jya}.


\section{Dressed-quark pressure}
We focus on 
the matter sector of QCD and hence begin with the quark gap equation:
\begin{subequations}
\label{eqGap}
\begin{eqnarray}
S(\vec{p},\tilde\omega_n)^{-1}
%
%
\label{eq:gap1}
&=&  i\vec{\gamma}\cdot\vec{p}
+ i\gamma_4 \tilde \omega_{n} + m
  + \Sigma(\vec{p}, \tilde\omega_{n}) \, ,\\
\nonumber
\Sigma(\vec{p},\tilde\omega_n) &=& T\sum_{l=-\infty}^\infty \! \int\frac{d^3{q}}{(2\pi)^3}\; {g^{2}} D_{\mu\nu} (\vec{p}-\vec{q}, \Omega_{nl}; T, \mu)\\
& & \times \frac{\lambda^a}{2} {\gamma_{\mu}} S(\vec{q},
\tilde\omega_{l}) \frac{\lambda^a}{2}
\Gamma_{\nu} (\vec{q}, \tilde\omega_{l},\vec{p},\tilde\omega_{n})\, ,
\label{eq:gap2}
\end{eqnarray}
\end{subequations}
where $m$ is the current-quark bare-mass; $\tilde\omega_{n}^{} = \omega_{n}^{} + i \mu$, $\omega_n=(2n+1)\pi T$ is the quark Matsubara frequency and $\mu$ is the quark chemical potential ($\mu_{B}^{} = 3 \mu$), $\Omega_{nl} = \omega_{n} - \omega_{l}$; $D_{\mu\nu}$ is the dressed-gluon propagator; and $\Gamma_{\nu}$ the dressed-quark-gluon vertex.  
%
%
%

The kernel of Eq.\,\eqref{eqGap} is determined by the quark-gluon vertex and the gluon propagator.  We use $\Gamma_\nu = \gamma_\nu$, which defines rainbow-ladder (RL) truncation, \emph{i.e}.\ the leading-order in the most widely used symmetry-preserving DSE approximation scheme \cite{Munczek:1994zz,Bender:1996bb}, which is accurate for ground-state light-quark hadrons \cite{Roberts:2000aa, Maris:2003vk, Chang:2011vu, Bashir:2012fs, Cloet:2013jya}.  It is appropriate here because whilst a certain class of vertex improvements can influence critical exponents associated with second order transitions, \emph{viz}.\ those leading to inclusion of long-range colour-singlet correlations \cite{Holl:1998qs}, no form available today alters the order of a transition or has a material impact on its location \cite{Qin:2010nq, Gao:2014rqa, Fischer:2014ata}.

The gluon propagator in Eq.\,\eqref{eqGap} has the general form
\begin{align}
& \tfrac{g^2}{8\pi^2} D_{\mu\nu}(\vec{k}, \Omega) = P_{\mu\nu}^{T} D_{T}(\vec{k}\,^2, \Omega^2) + P_{\mu\nu}^{L} D_{L}(\vec{k}\,^2, \Omega^2)\,,
\end{align}
where $P_{\mu\nu}^{T,L}$ are, respectively, $\vec{k}$ transverse and longitudinal projection operators, and  $P_{\mu\nu}^{T}+P_{\mu\nu}^{L}=\delta_{\mu\nu}-k_\mu k_\nu/k^2$; $D_{T} =\mathcal{D}({s_{\Omega}},0)$,
$D_{L} =\mathcal{D}({s_{\Omega}},{m_{g}^{2}})$,
where $m_{g}^{2}=(16/5)(T^2+6\mu^2/[5\pi^2])$ describes a gluon screening mass, the value of which is determined from leading-order perturbative QCD \cite{Haque:2012my}; and ($s_{\Omega}^{} = \Omega^2 + \vec{k}\,^2 + m^{2}_{g}$) \cite{Qin:2011dd}
\begin{equation} \label{eq:DGP-QCmodel}
\mathcal{D}({s_{\Omega}}, {m_{g}^{2}}) =
\frac{D}{\omega^{4}} e^{-{s_{\Omega}^{}}/\omega^{2}}
 + \frac{ {\gamma_{m}}{\cal F}(s_{\Omega}^{})}{{\ln}[ \tau \! + \! (1 \! + \!
{s_{\Omega}^{}}/{\Lambda_{\text{QCD}}^{2}} ) ^{2} ] } \,
 \, ,
\end{equation}
with $z {\cal F}(z) = (1-e^{-z/4 \omega^{2}})$, $\omega=0.5\,$GeV, $\tau=e^2-1$, $\gamma_m=12/25$,  $\Lambda_{\text{QCD}}=0.234\,$GeV.
%
$\mathcal{D}({s_{\Omega}}, {m_{g}^{2}})$ is shape-consistent with solutions of in-vacuum gap equations \cite{Binosi:2014aea}; and with $D\omega = (0.8\,{\rm GeV})^3$ and a renormalisation-group-invariant current-quark mass $\hat m_{u,d} = 6\,$MeV, the solutions of Eq.\,\eqref{eqGap} support a reliable in-vacuum description of ground-state hadrons in RL truncation \cite{Qin:2011dd}.  (\emph{N.B}.\ By using Landau gauge we minimise sensitvity to truncation-induced violations of gauge covariance \cite{Bashir:2011dp}.)

The in-medium extension of the gap equation's kernel, Eq.\,\eqref{eq:DGP-QCmodel}, preserves agreement with QCD at large momenta.  However, in assuming that $D$ is $(\mu,T)$-independent it overlooks screening of the interaction's infrared strength.  We remedy that by writing \cite{Qin:2010nq, Gao:2014rqa}
\begin{eqnarray}
D(T,\mu)=D \left\{
\begin{array}{ll}
\displaystyle
1 \,, &   T<T_{\text{p}} \, ,  \\
\displaystyle
\frac{a}{b(\mu)+ \ln[\tilde T/\Lambda_{QCD}]\rule{0ex}{2.5ex}}\,, &  T \ge
T_{\text{p}} \, ,
\end{array}
\right.  \label{DTfunction}
\end{eqnarray}
where $\tilde T^2=T^2+6\mu^2/[5\pi^2]$ and $T_{\rm p}$ marks the onset of  thermal screening.  With $T_{\rm p}=T_{c}$, the critical temperature for chiral symmetry restoration, then $a=0.029$, $b=0.47$ yield a dressed-quark thermal mass $m_T = 0.8\,T$ at $T=2T_c$, in agreement with lQCD simulations \cite{Karsch:2009tp}.  Naturally, $T_c = T_c(\mu)$: we set $T_{\rm p}(\mu)=T_{c}(\mu)$ at $\mu\neq 0$, ensuring $D(T_{c}(\mu),\mu )=D$ by evolving the value of $b(\mu)$.

\begin{figure}[t]
\centerline{\includegraphics[width=0.40\textwidth]{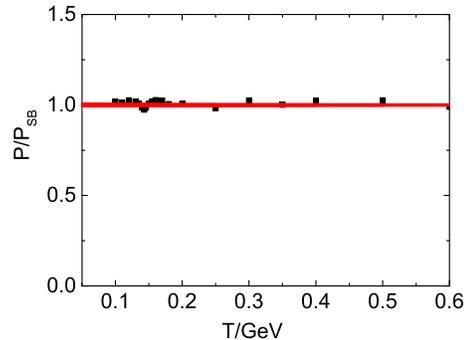}}
\caption{\label{freepressure}
Points -- Numerical result for the free-quark pressure obtained as described in connection with Eq.\,\eqref{eq:Kapusta}; and solid (red) curve -- analytical result.}
\end{figure}

In RL truncation and stationary phase approximation, the dressed-quark pressure density is  \cite{Haymaker:1990vm}:
\begin{equation}
\label{eq:pressure}
P[S]=T \ln{Z}=-T ( {\rm Tr}\ln{[T S]}+\tfrac{1}{2}Tr[\Sigma S] )\, ,
\end{equation}
where Eq.\,\eqref{eqGap} determines $S$, $\Sigma$.  At each $(\mu,T)$, Eq.\,\eqref{eq:pressure} possesses the same ultraviolet divergence, which may be eliminated by subtracting the $\mu=0=T$ result.  The subtraction can be accomplished by recalling that for any function $f(w)$, compatible with a physical system, \cite{Kapusta:1989}:
\begin{align}
\nonumber
 & 2\pi i T \! \sum^{\infty}_{n=-\infty} f(i\omega_{n}^{}+\mu)
= \int_{u_0^\ast}^{u_0} \!dw\, f(w) \\
&
-\int_{u_\mu^\ast+\eta}^{u_\mu+\eta} \! 
\frac{dw\, f(w)}{{\rm e}^{(w-\mu)/T}+1}
-\int_{u_\mu^\ast-\eta}^{u_\mu-\eta} \!  
\frac{dw\,f(w)}{{\rm e}^{-(w-\mu)/T}+1}
\,,
\label{eq:Kapusta} 
\end{align}
where we have omitted a $T$-independent term that is zero so long as $\mu<\mu_c(T)$, and $u_\mu=i\Lambda+\mu$ with $\Lambda\to\infty$, $\eta\to 0$.  The first term on the right-hand-side of Eq.\,\eqref{eq:Kapusta} is
responsible for the divergence we wish to eliminate.  The physical piece of the dressed-quark pressure can thus be calculated as the difference between the two terms in the first line of Eq.\,\eqref{eq:Kapusta}, with $f(w)$ computed via the functional expression in Eq.\,\eqref{eq:pressure}.
Practically, one proceeds as follows:
solve the gap equation at a given $(\mu,T)$-pair for a large number of Matsubara frequencies, characterised by $n_m$;
at each $\mu$, obtain smooth interpolations in ``$w$'' for the scalar functions obtained thereby, Eq.\,\eqref{eqGap};
evaluate $P[S]$ with these inputs;
and then compute the difference between the sum and integral, verifying that it is insensitive to the interpolation procedure and choice of $n_m$.
That this procedure can work effectively is illustrated in Fig.\,\ref{freepressure}, which displays a comparison of our numerical result for the free-quark pressure with the analytic form.

Our approximation for the dressed-quark contribution to the QCD EoS is now defined.
Using the pressure, one can define the \emph{trace} \emph{anomaly}: ${\mathcal I}=\varepsilon - 3P$, where the energy density $\varepsilon = -P+ T s$ and $s(T)=\partial P[S]/\partial T$ is the entropy density.  Notably, ${\mathcal I}\equiv 0$ for an noninteracting ultrarelativistic gas, which is described by $P_{\rm SB}\propto T^4$; and hence ${\mathcal I}$ is a measure of the interaction energy stored in the system.
Now, since the confined dressed-quark contribution to the physical pressure must vanish on $T \simeq 0$, it follows that ${\mathcal I}$ exhibits a maximum at some $T_M$, the value of which serves to define a useful reference temperature.


\section{Phase diagram and thermodynamics}
We employ the DSEs because they possess the capacity to study confinement and DCSB simultaneously in the continuum \cite{McLerran:2007qj}.  Within this framework the $(\mu\neq 0,T\neq 0)$ EoS has only been computed using a very simple description of QCD's gauge sector \cite{Blaschke:1997bj}.  Owing to the importance of the EoS in developing a complete picture of the Standard Model, it is imperative to do better.

\begin{figure}[t]
\centerline{\includegraphics[width=0.35\textwidth]{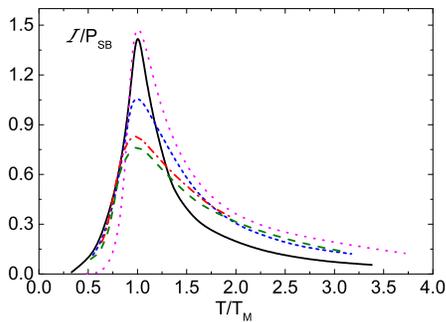}}
\caption{\label{fig:trace}
Trace anomaly (normalised, $\mu=0$): ${\mathcal I}/P_{\rm SB}$.  $P_{\rm SB}$ is the pressure of a noninteracting ultrarelativistic gas containing the number of gluons and quarks appropriate to the calculation.
Solid (black) curve -- dressed-quark contribution, computed via Eq.\,\eqref{eq:Kapusta}.
For comparison, lQCD results obtained using various discretisation schemes.
Complete pressure:
short-dashed (blue) \cite{Huovinen:2009yb};
long-dashed (green) \cite{Borsanyi:2012crS};
and dot-dashed (red)  \cite{Bazavov:2014pvz}.
%
%
Gluon-only contribution: dotted (pink) curve \cite{Boyd:1996bx}.}
\end{figure}

We depict the $\mu=0$ trace anomaly in Fig.\,\ref{fig:trace}.  In order to facilitate comparisons between the profiles obtained in different analyses, the temperature is expressed in units of the appropriate value of $T_M$.  The value of $T_M$ exhibits modest variation between the calculations: in our case $T_M=0.14\,$GeV; which is roughly 30\% smaller than found in modern lQCD studies owing to our omission of the gluon contribution.
%
%
Additionally, each computation in Fig.\,\ref{fig:trace} is normalised by the appropriate form of $P_{\rm SB}$: represented in this way, there is qualitative agreement between all results.  Our prediction describes the quark-only contribution to ${\mathcal I}$.  It matches the lQCD results in shape and order of magnitude.  These observations indicate both that the gluon and quark contributions to the total interaction energy behave similarly and that they are commensurate in size when measured against their respective asymptotic contributions to the pressure.

\begin{figure}[t]
\centerline{\includegraphics[width=0.35\textwidth]{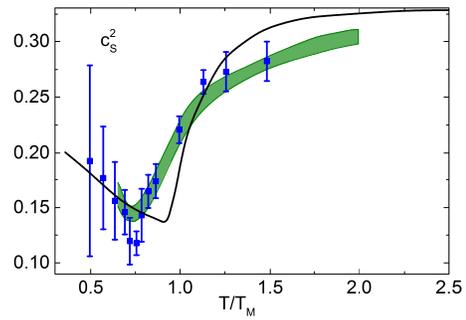}}
\caption{Temperature dependence of $c_s^2$, where $c_s$ is the sound velocity in the system.
Solid (black) curve -- our result, computed from the dressed-quark pressure;
points (blue) -- lQCD results from Ref.\,\cite{Borsanyi:2012crS};
and band (green) -- lQCD results from Ref.\,\cite{Bazavov:2014pvz}.}
\label{fig:soundspeed}
\end{figure}

The ``speed of sound'' in the system is obtained from $c_{s}^{2} = \partial P/\partial \varepsilon $.  It is a crucial factor in determining the flow of material in the system, \emph{i.e}.\ its transport properties.  As evident in Fig.\,\ref{fig:soundspeed}, our prediction for the sound velocity in the dressed-quark subcomponent is similar to results obtained in lQCD for the velocity in the complete system.

\begin{figure}[t]
\begin{minipage}[t]{0.5\textwidth}
\begin{minipage}{0.49\textwidth}
\centerline{\includegraphics[clip,width=\textwidth]{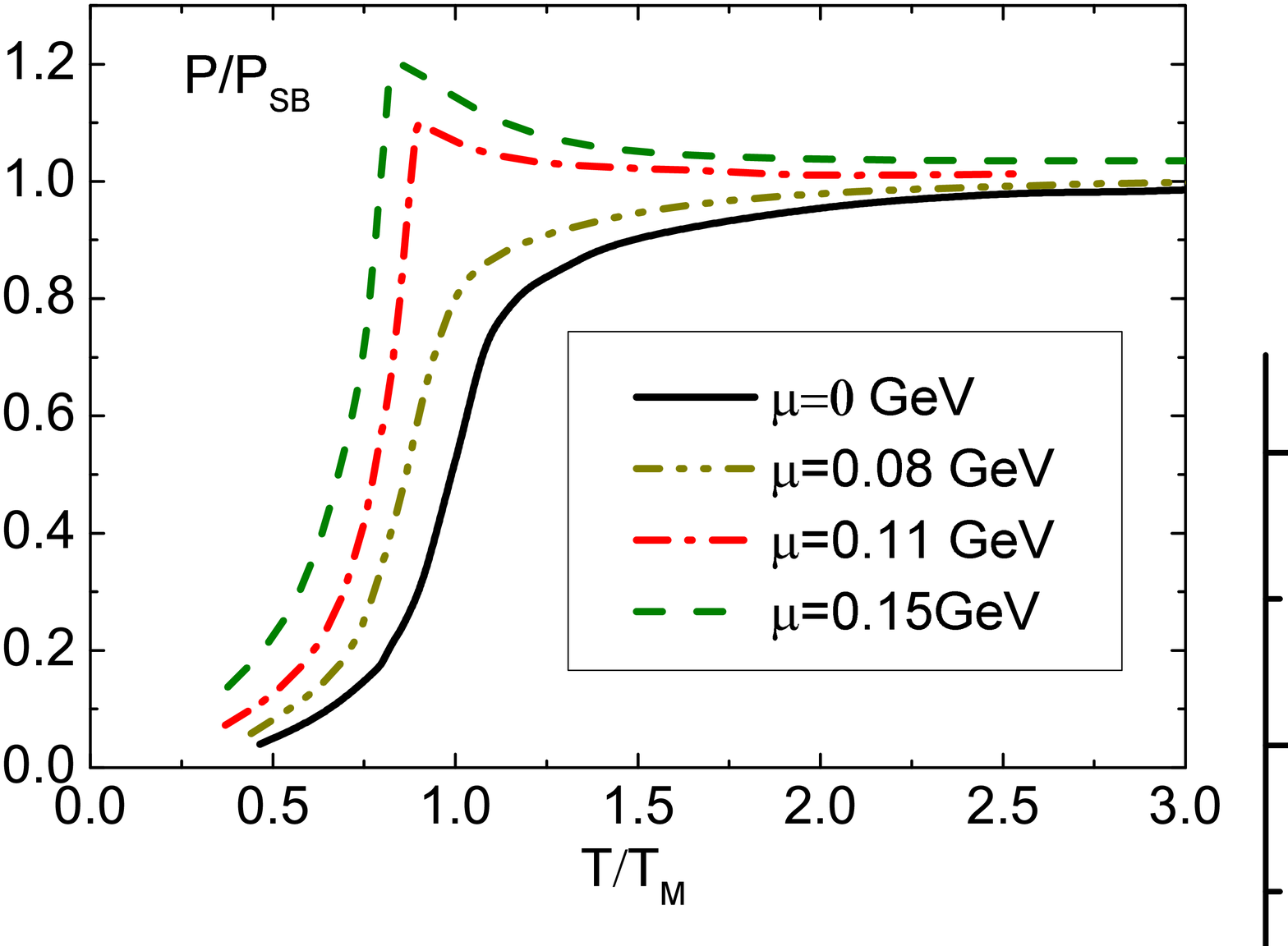}}
\end{minipage}
\begin{minipage}{0.49\textwidth}
\hspace*{-2ex}{\includegraphics[width=\textwidth]{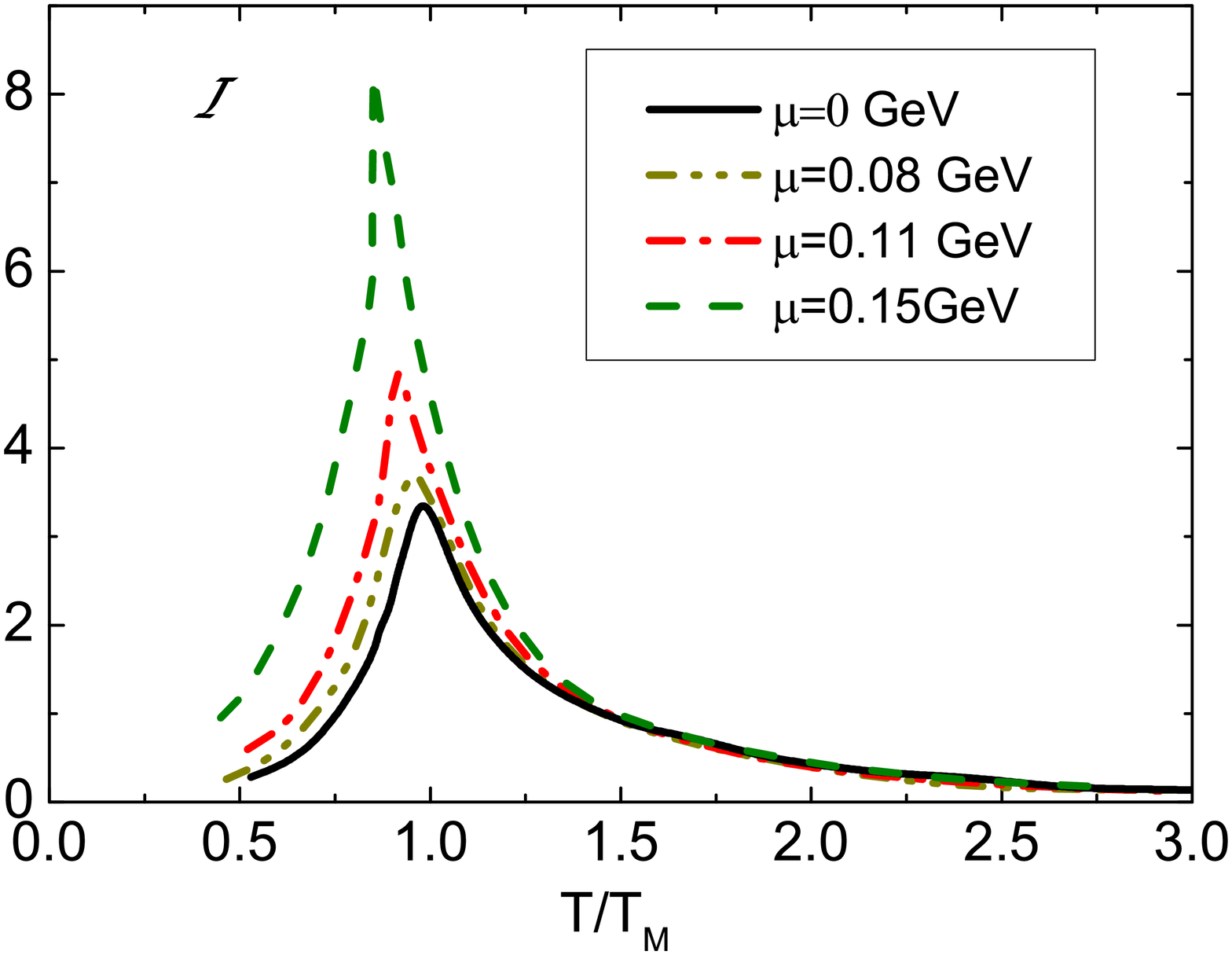}}
\end{minipage}
\end{minipage}
\begin{minipage}[t]{0.5\textwidth}
\begin{minipage}{0.49\textwidth}
\centerline{\includegraphics[clip,width=\textwidth]{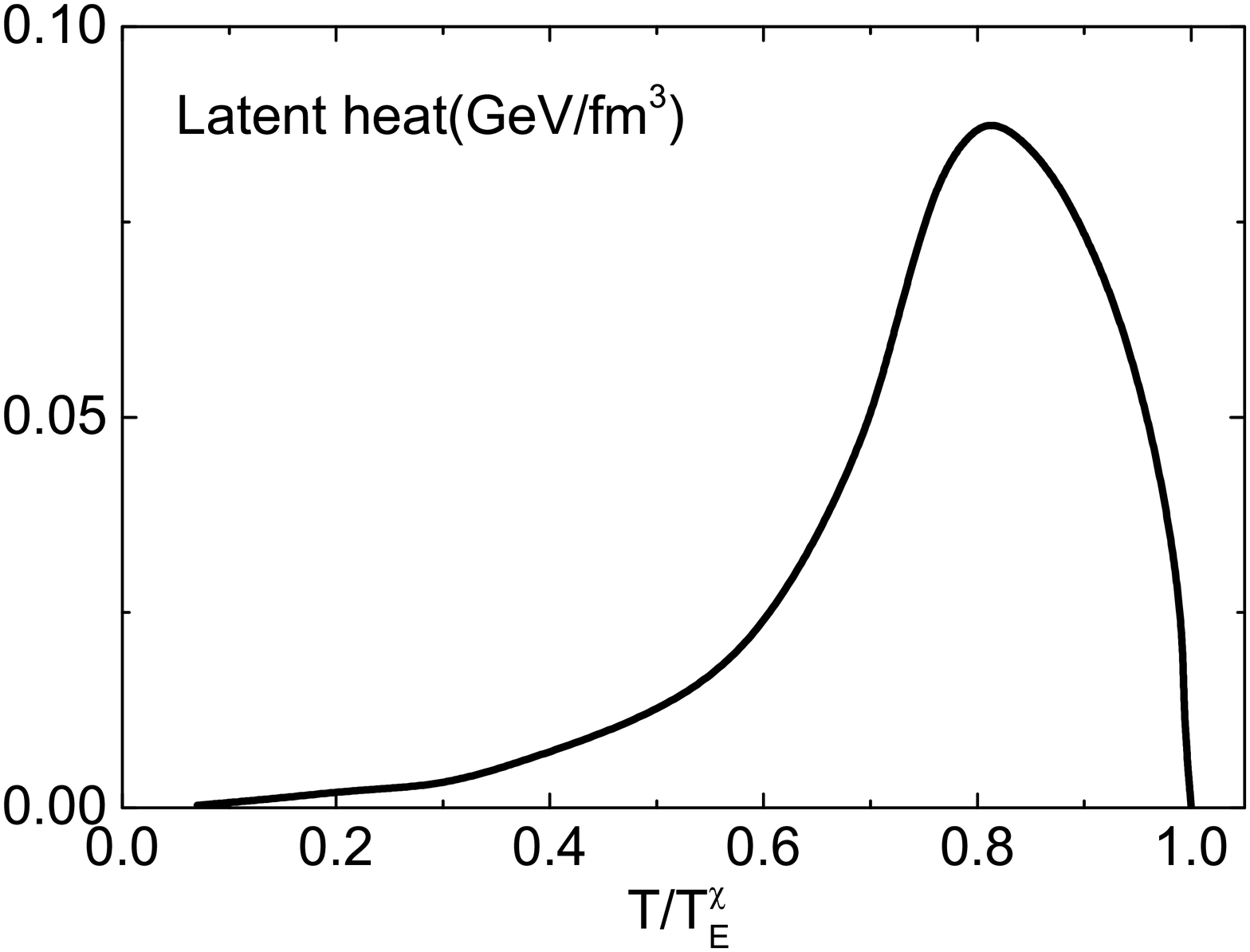}}
\end{minipage}
\begin{minipage}{0.49\textwidth}
\hspace*{-3ex}\includegraphics[width=\textwidth]{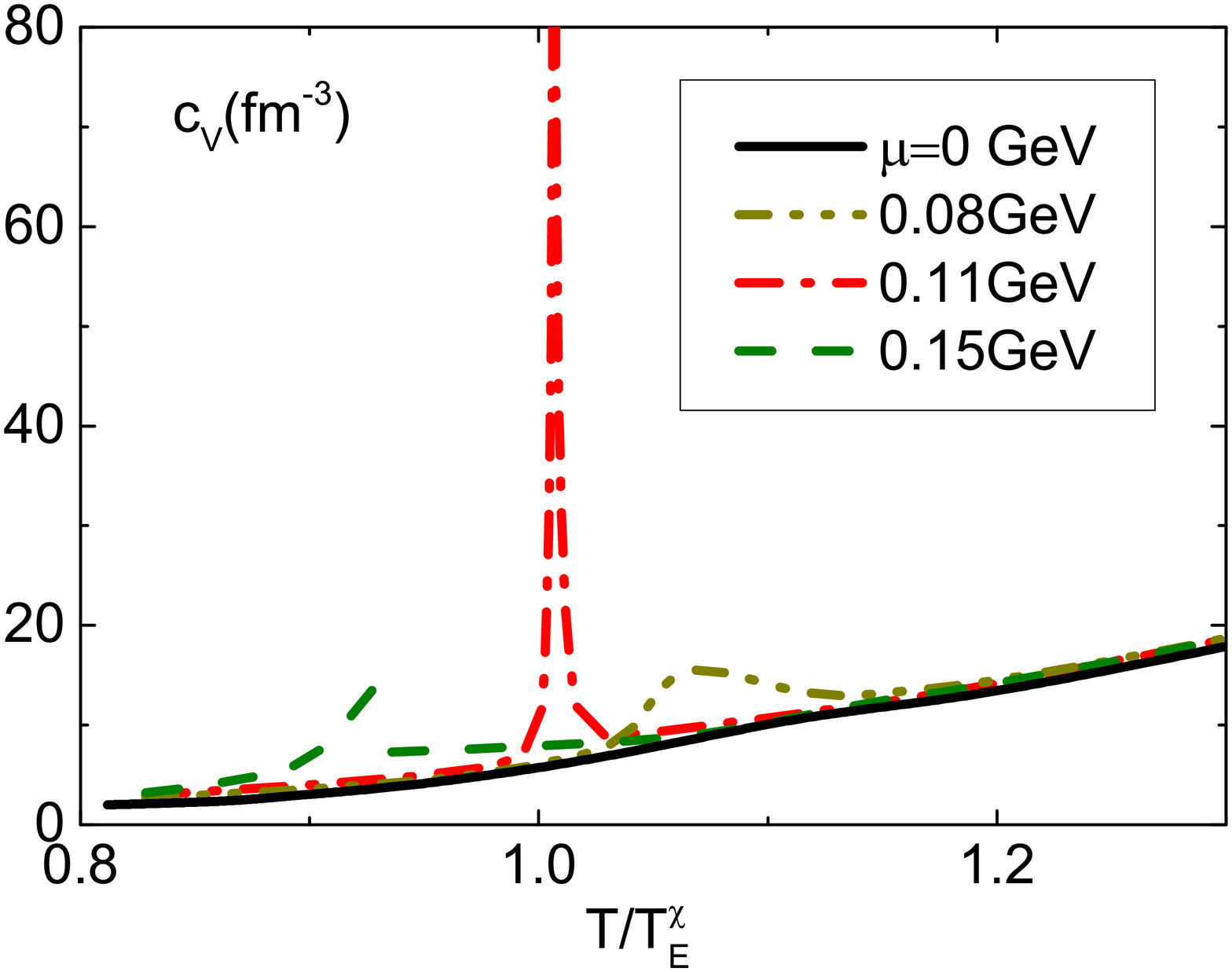}
\end{minipage}
\end{minipage}
\caption{\emph{Upper panels}: Temperature dependence of the dressed-quark pressure (\emph{left}) and trace anomaly (\emph{right}) at several chemical potentials.
\emph{Lower left panel}: latent heat of the chiral symmetry restoring transition.
\emph{Lower right panel}: heat capacity of the system's dressed-quark subcomponent.
\label{fig:full}}
\end{figure}

A feature which distinguishes our framework from lQCD is its ability to treat $\mu>0$ without further approximation.  Thus, in Fig.\,\ref{fig:full} we display results for the dressed-quark pressure and trace anomaly at a range of values of $\mu>0$: whilst increasing $\mu$ produces an increase in the pressure at all values of temperature, it only materially increases the interaction energy on $T<T_M$.  Similar behaviour is seen in lQCD estimates for the $\mu$-dependence of these quantities, obtained using various extrapolation schemes \cite{Fodor:2002km, Allton:2003vx, DeTar:2010xm, Borsanyi:2012crS}.
In detail, $P$ is a monotonic function of $T$ for small $\mu$, and $P$ and ${\mathcal I}$ are smooth; but qualitative changes occur at $\mu=\mu_p=0.106\,$GeV: a peak appears in $P$ and that in ${\mathcal I}$ becomes sharper.  Both functions remain smooth, however, until $\mu=0.111\,$GeV, whereat the first derivative of each diverges at $T=0.128\,$GeV, signalling that the transition has become first-order.  This effect locates the critical endpoint (CEP) for the chiral symmetry restoring transition at $(\mu_E^\chi = 0.111,T_E^\chi=0.128)\,$GeV.
%
The behaviour of both $P$ and ${\mathcal I}$ on $\mu/T \simeq 0$ is consistent with hard thermal loop perturbation theory \cite{Haque:2012my}.

We now draw the diagrams associated with QCD's phase transitions as determined from the dressed-quark pressure.  Chiral symmetry restoration is straightforward.  It may be charted via the $(\mu,T)$-dependence of the chiral condensate \cite{Brodsky:2010xf, Chang:2011mu, Brodsky:2012ku}; but we prefer a method \cite{Qin:2010nq} based on the chiral susceptibility, $\chi(\mu,T)$.

Before discussing deconfinement, however, one must have a definition of colour confinement.  We consider confinement as a violation of reflection positivity by coloured Schwinger functions.  This associates it with dynamically-driven changes in the analytic structure of QCD's propagators and vertices \cite{Gribov:1999ui, Munczek:1983dx, Stingl:1983pt, Cahill:1988zi, Krein:1990sf, Hawes:1993ef} that occur because both gluons and quarks acquire running mass distributions, which are large at infrared momenta.  This leads to the emergence of a length-scale $\varsigma \approx 0.5\,$fm, whose existence and size is evident in all continuum and lQCD studies of dressed-gluons and -quarks: $\varsigma$ characterises the material change in their analytic structure \cite{Bhagwat:2003vw, Bowman:2007du, Karsch:2009tp, Strauss:2012dg}.  From this perspective, deconfinement occurs when $\varsigma\to 0$ and reflection positivity is thus recovered.  This criterion has been used effectively in dense-hot QCD (\emph{e.g}.\ Refs.\,\cite{Bender:1996bm, Bender:1997jf, Blaschke:1997bj, Mueller:2010ah, Qin:2010pc, Qin:2013ufa, Gao:2014rqa}; and we employ it herein, following a local implementation elucidated in Refs.\,\cite{Roberts:2007ji, Bashir:2008fk, Bashir:2009fv}.

\begin{figure}[t]
\centerline{\includegraphics[width=0.37\textwidth]{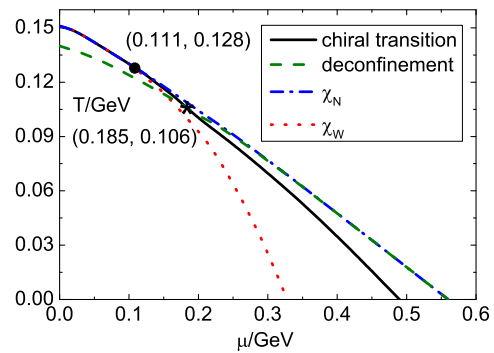}}
\caption{\label{fig:phase}
Deconfinement and chiral symmetry restoration phase boundaries, computed via pressure in Eq.\,\eqref{eq:Kapusta}.
Curves:
solid (black) -- chiral transition, with DCSB favoured below the curve;
dashed (green) -- deconfinement transition, with dressed-quarks confined below the curve;
dot-dashed (blue) -- Nambu phase chiral susceptibility, $\chi_N$: positive below the curve and zero above;
dotted (red) -- Wigner phase chiral susceptibility, $\chi_W$: positive \emph{above} the curve and negative below.
CEPs: chiral -- filled circle; and deconfinement -- asterisk.
}
\end{figure}

We display the phase diagram computed from the dressed-quark pressure in Fig.\,\ref{fig:phase}.  Comparison with Fig.\,3 in Ref.\,\cite{Qin:2010nq} shows that our improved DSE kernel, which agrees with the one-loop QCD renormalisation group, does not qualitatively alter the phase diagram.  The solid curve in Fig.\,\ref{fig:phase} is the locus of transition: the Nambu phase is energetically favoured for those values of $(\mu,T)$ that lie within the domain bounded by the axes and this curve:
 \begin{equation}
 \label{P0locus}
T_c^P(z=\mu/T_{c0}^P)= T_{c0}^P  \frac{1+ 0.52 z^2 - 0.058 z^4}{1+0.91 z^2}\, .
\end{equation}
The $\mu=0$ pseudo-critical temperature associated with the chiral transition is $T_{c0}^P=0.15\,$GeV.  For comparison, a lQCD estimate of the critical temperature for chiral symmetry restoration in QCD with two light flavours and a physical strange quark mass is $T_c=0.15\pm 0.01\,$GeV \cite{Bazavov:2011nk}.  Within a factor of two, the $z=0$ slope of $T_c^P(z)/T_{c0}^P$ from Eq.\,\eqref{P0locus} agrees with estimates from lQCD \cite{Borsanyi:2012crS}.

%

With $(\mu,T)$ increasing from the origin, the dot-dashed curve in Fig.\,\ref{fig:phase} bounds the domain of positive Nambu-phase chiral susceptibility.
 The dotted curve, on the other hand, marks the line whereat the Wigner-phase chiral susceptibility switches from negative to positive.
 These curves coincide with the transition locus, Eq.\,\eqref{P0locus}, up to the chiral transition's CEP:
 \begin{equation}
 \mbox{CEP}_\chi=(\mu_E^\chi = 0.111,T_E^\chi=0.128)\,\mbox{GeV}\,,
 \end{equation}
 which confirms the result obtained in connection with Fig.\,\ref{fig:full}, top panels; and the computed ratio $\mu_E^\chi/T_E^\chi = 0.87$ is commensurate with those in Refs.\,\cite{Fodor:2004nz, Schmidt:2008ev, Qin:2010nq, Fischer:2012vc}.
 The chiral crossover becomes a first-order transition at CEP$_\chi$: the Nambu and Wigner phases coexist, with the Nambu phase dominant below the transition locus and the Wigner phase dominant otherwise.

The dashed curve marks the boundary for the deconfinement transition, which is second-order until
\begin{equation}
\mbox{CEP}_\varsigma=(\mu_E^\varsigma = 0.185, T_E^\varsigma=0.106)\,\mbox{GeV}\,,
\end{equation}
$\mu_E^\varsigma/T_E^\varsigma=1.75$.  Wigner phase-domains exhibit neither confinement nor DCSB at any values of $(\mu,T)$.
 In the chiral limit the solid and dashed curves coincide and CEP$_\chi=\;$CEP$_\varsigma$: chiral symmetry restoration and deconfinement are coincident; but the dislocation at $\hat m\neq 0$ entails the existence of a small domain wherein quarks are deconfined but chiral symmetry is broken.  This is the set $\{\mu,T\}$ enclosed within the dot-dashed and dashed curves.  Pockets of Nambu phase material in this subset of the phase coexistence region possess that character.
 %
 %
In the domain of $(\mu,T)$ enclosed between the axes and the dashed and dotted curves, the system is in a pure phase with confinement and DCSB, whereas chiral symmetry is restored and quarks are deconfined in the domain above the dot-dashed curve.

We can now compute the heat-capacity density of the dressed-quark system, $c_V =\partial \varepsilon /\partial T$, and the latent-heat density of transition:
$L = T\Delta s=\Delta \varepsilon-\mu\Delta \rho$, $\rho(T)=\partial P/\partial \mu$ is the quark number density and $\Delta_F$ is the difference between the quantity $F$ in the two distinct phases, which are here the Nambu (chirally asymmetric) and Wigner (chiral symmetry restored) phases.  We depict $L$, computed along the phase boundary, which is the trajectory in Eq.\,\eqref{P0locus}, in the lower-left panel of Fig.\,\ref{fig:full}.  Naturally, $L=0$ for $T\geq T_E^\chi$ because the transition is no longer first order.  Otherwise our prediction is qualitatively consistent in shape with what may be inferred from lQCD simulations \cite{Borsanyi:2012crS}.

%
%

The lower-right panel of Fig.\,\ref{fig:full} displays our prediction for $c_V$: it diverges as $(\mu,T)\to\,$CEP$_\chi$.  Actually, in the neighbourhood of the CEP one may write \cite{Hatta:2002sj}: $c_V \propto |g-g_E^\chi|^{-\epsilon}$, where $g=\mu$, $T$; and the quark number susceptibility $\chi = \partial \rho/\partial \mu$ behaves in the same fashion.
Analysing our results, we find $\epsilon = 0.67 \pm 0.02$ in both cases.  These are the critical exponents of a mean-field transition, which is the nature of RL truncation.


\section{Summary}
We introduced a practical procedure for computing the $(\mu,T)$-dependent dressed-quark pressure in continuum QCD, which we illustrated using a gap equation whose solutions are key to a successful description of the properties of ground-state hadrons in-vacuum.
Without further approximation, the associated richly-structured phase diagram in the $(\mu,T)$-plane was computed.  We drew the transition lines for deconfinement and chiral symmetry restoration, and confirmed that these transition are identical in the chiral limit.
Likewise, we calculated the speed of sound in the system and provided the $(\mu,T)$-dependence of the trace anomaly, latent-heat and heat-capacity densities.
Where comparisons are possible, our predictions are consistent with results from lattice QCD.  Notably, predictions obtained from the dressed-quark pressure are qualitatively equivalent to those computed using the complete pressure, which suggests that the dressed-quark pressure alone can be used as a practical guide to some of QCD's thermal properties.
No material improvement over our results can be envisaged in a continuum analysis before a symmetry-preserving kernel including long range correlations is derived for the gap equation.  Our method for computing the pressure will also be applicable then.


\acknowledgments

%
Work supported by:
National Natural Science Foundation of China under Contract Nos.\ 11435001 and 11175004;
the National Key Basic Research Program of China under Contract Nos. G2013CB834400 and 2015CB956900;
the Office of the Director at Argonne National Laboratory through the Named Postdoctoral Fellowship Program;
and the U.S.\ Department of Energy, Office of Science, Office of Nuclear Physics, under contract no.~DE-AC02-06CH11357.
%


\end{document}